\documentclass{ifacconf}

\usepackage{graphicx}           
\usepackage{tikz}               
\usepackage{pgfplots}           
\usepackage{pgfplotstable}      
\usepackage{fontawesome5}       
\usepackage[utf8]{inputenc}     
\usepackage{comment}            
\usepackage{xcolor}             
\usepackage{amsfonts}           
\usepackage{amsmath}            
\usepackage{mathtools}          
\usepackage{amssymb}            
\usepackage{bm}                 
\usepackage{booktabs}           
\usepackage{tabularx}           
\usepackage{multirow}           
\usepackage{makecell}           
\usepackage{setspace}           
\usepackage{colortbl}           
\usepackage{times}              
\usepackage{natbib}             
\usepackage{enumitem}           
\usepackage{cases}              
\usepackage{etoolbox}           
\usepackage[hyphens]{url}       

\allowdisplaybreaks 

\usetikzlibrary{calc,arrows.meta,positioning,shapes.geometric,fit,external}
\usepgfplotslibrary{groupplots,statistics,external}
\pgfplotsset{compat=1.18}

\definecolor{meta_color}{RGB}{115, 0, 140}       
\definecolor{meta_sim_color}{RGB}{115, 0, 140}   
\definecolor{meta_step_color}{RGB}{186, 85, 211} 
\definecolor{indiv_color}{RGB}{44, 160, 44}      
\definecolor{narx_color}{RGB}{190, 20, 20}       
\definecolor{narx_sim_color}{RGB}{190, 20, 20}    
\definecolor{narx_step_color}{RGB}{240, 115, 60}  
\definecolor{real_color}{RGB}{0, 0, 0}           

\newcommand{\paragraphfont}[1]{\textsc{#1}.} 

\newtheorem{remark}{Remark}

\newcommand{\R}{\mathbb{R}}
\newcommand{\N}{\mathbb{N}}
\newcommand{\Z}{\mathbb{Z}}
\newcommand{\Zpos}{\Z^+}
\newcommand{\iset}[2]{\mathbb{I}_{#1}^{#2}}
\newcommand{\trans}{\top}

\DeclarePairedDelimiter\pr{\lparen}{\rparen}
\DeclarePairedDelimiter\ps{\lbrack}{\rbrack}
\DeclarePairedDelimiter\pc{\lbrace}{\rbrace}

\DeclarePairedDelimiterX\norm[1]{\lvert}{\rvert}{%
    \ifblank{#1}{\,\cdot\,}{#1}%
}
\DeclarePairedDelimiter\abs{\lvert}{\rvert}

\DeclarePairedDelimiterXPP\set[3]{}{\lbrace}{\rbrace}{%
    \ifblank{#2}{}{_{#2}}\ifblank{#3}{}{^{#3}}%
}{%
    #1%
}

\newcommand{\context}{\gamma}
\newcommand{\bcontext}{\bm{\context}}
\newcommand{\embedding}{\varepsilon}
\newcommand{\bembedding}{\bm{\embedding}}
\newcommand{\information}{\iota}
\newcommand{\binformation}{\bm{\information}}
\newcommand{\outputvec}{\omega}
\newcommand{\boutputvec}{\bm{\outputvec}}
\newcommand{\parameters}{\theta}
\newcommand{\bparameters}{\bm{\parameters}}
\newcommand{\bParameters}{\bm{\Theta}}
\newcommand{\ds}{\mathcal{D}}
\newcommand{\loss}[1]{\mathcal{L}_{#1}}
\newcommand{\lossw}[1]{\alpha_{#1}}

\newcommand{\settablestretch}{\setstretch{1.05}}

\begin{document}
\begin{frontmatter}

\title{Transfer Meta-Learning for State Space Autoencoders} 

\thanks[corresp]{Corresponding author: \texttt{stefano.decarli@unibg.it}. This work has been submitted to IFAC for possible publication.}

\author[unibg]{Stefano De Carli\thanksref{corresp}}
\author[gssi]{\quad Daniele Masti}
\author[unibg]{\quad Davide Previtali}
\author[unibg]{\quad Mirko Mazzoleni}

\address[unibg]{Department of Management, Information and Production Engineering, University of Bergamo, Italy}
\address[gssi]{Gran Sasso Science Institute (GSSI), L'Aquila, Italy}
%
\begin{abstract}
    \label{s:abstract}
    Nonlinear system identification often requires costly experimental campaigns and separate model fitting for each different instance (target system) of a given system family.
    To reduce this burden, meta-learning approaches reuse information acquired across related systems.
    In particular, in-context meta-learning enables zero-shot prediction from a brief target system context (i.e., a short input--output trajectory), but generally yields large black-box sequence predictors without exposing an explicit state space model.
    Conversely, compact structured state space models are suitable for simulation, estimation, and control, but are usually fitted or adapted separately for each target system.
    To bridge this gap, we introduce a transfer meta-learning extension of state space autoencoders.
    The method reuses a pretrained encoder, whose weights remain fixed, to extract a compact representation of the target system from its context.
    An affine parameterization maps this representation to the parameters of a compact neural state space model without target-specific adaptation; the resulting model contains orders of magnitude fewer parameters than the reference in-context predictors from the literature.
    We evaluate zero-shot prediction and simulation performance on held-out system instances from Buck--Boost converter and Bouc--Wen oscillator system families, comparing performance against individually fitted baseline models.
\end{abstract}


\end{frontmatter}

\section{Introduction}
\label{s:introduction}
%
In conventional system identification workflows, a dedicated model is estimated from data collected on the system under consideration, with limited to no reuse of data acquired from related systems~\citep{chakrabartyMetaLearningNeuralStateSpace2023}.
This approach becomes especially impractical when commissioning large populations of similar devices, as conducting informative excitation experiments on every single unit is costly, time-consuming, and potentially intrusive.
A promising solution to this hurdle is offered by \emph{few-shot learning} techniques, where knowledge acquired across related system instances is leveraged to initialize a model, which is subsequently adapted using only limited data collected from the unseen target system.
Within this paradigm, two primary strategies exist: \emph{transfer learning}, which repurposes features or weights pretrained on possibly different source domains to accelerate target modeling, and \emph{meta-learning}, which explicitly optimizes a shared model structure alongside a dedicated mechanism for rapid target adaptation~\citep{finnModelAgnosticMetaLearningFast2017, kirchmeyerGeneralizingNewPhysical2022, blankeInterpretableMetaLearningPhysical2024}.
In neural network-based state space modeling, meta-learning has been demonstrated to significantly outperform conventional model-level transfer learning (i.e., fine-tuning the weights of a model pretrained on a single related system) in scarce data regimes~\citep{chakrabartyMetaLearningNeuralStateSpace2023}.

Most established meta-learning algorithms rely on \emph{optimization-based adaptation}, such as model-agnostic meta-learning and its variants~\citep{finnModelAgnosticMetaLearningFast2017, chakrabartyMetaLearningNeuralStateSpace2023}.
Although effective, these methods still require a target-specific training phase at deployment time.
More recently, several \emph{in-context meta-learning} approaches have emerged as an appealing training-free alternative, where a sequence model directly processes a short input--output record, i.e., the so-called \emph{context}, and predicts the target dynamics in a purely feedforward manner~\citep{forgioneSystemModelsClass2023, rufoloEnhancedTransformerArchitecture2025}.
Since these methods infer target dynamics from context through feedforward evaluation without target-specific adaptation, we refer to these in-context approaches as \emph{zero-shot}.

Despite their flexibility, current in-context system identification approaches leave open an important \emph{gap between rapid zero-shot prediction and usability of the learned models within classical model-based engineering tools}.
Indeed, many of them rely on large sequence models, such as Transformers or deep state space sequence models, whose large parameter count and memory footprint complicate embedded deployment or real-time execution on resource-constrained platforms~\citep{forgioneSystemModelsClass2023, rufoloEnhancedTransformerArchitecture2025}.
Second, these sequence models predict the input--output behavior of an unseen system without necessarily exposing useful state dynamics or a compact state space representation suitable for downstream estimation and control~\citep{forgioneSystemModelsClass2023, rufoloEnhancedTransformerArchitecture2025}.
Furthermore, in-context meta-models are inherently tied to the specific system class observed during their initial training.
When faced with fundamentally different system classes, existing approaches require computationally demanding retraining from scratch or tuning of the entire architecture~\citep{pigaAdaptationIncontextLearners2024}.
Transferring representations learned by a pretrained meta-model across distinct system classes without updating its parameters therefore remains largely unexplored in system identification.

This motivates a twofold research question:
\begin{description}[nosep, leftmargin=2.0em]
    \item[\textbf{RQ1:}] \emph{Can a pretrained component from an in-context system identification framework extract informative representations when transferred across distinct system classes?}
    \item[\textbf{RQ2:}] \emph{Can these representations be used to instantiate a compact state space model suitable for state estimation and model-based control, without target-specific optimization?}
\end{description}

We address both questions by introducing \emph{Transfer-Meta State Space AutoEncoders (Transfer-Meta SSAEs)}.
Specifically, we extend the SSAE identification framework of~\citet{mastiLearningNonlinearState2021} to a transfer meta-learning setting by generating its model parameters from a brief measured input--output trajectory.
We reuse a pretrained encoder from~\citet{rufoloEnhancedTransformerArchitecture2025}, keeping its weights fixed, and learning to aggregate its representations into a compact characterization of the target system.
Inspired by the meta-learning approach of~\citet{forgioneManifoldMetalearningReducedcomplexity2025}, an affine mapping directly computes the SSAE model parameters from this representation.
Our approach is therefore zero-shot: we obtain the target-model parameters through feedforward evaluation, without target-specific adaptation or optimization.
After this instantiation step, we use the resulting SSAE for prediction and simulation, with fixed model parameters and without further context evaluations.
We evaluate our approach in a data-scarce regime across two parameterized system families, namely Buck--Boost converters and Bouc--Wen oscillators.
For each family, the model is evaluated on held-out parameterizations and compared with individually fitted SSAEs and a nonlinear input--output baseline in prediction and simulation.

\paragraphfont{Organization}
The remainder of this paper is organized as follows.
Section~\ref{s:problem_statement} states the identification problem for a family of nonlinear systems.
Section~\ref{s:meta_ssae} introduces the SSAE model, develops its zero-shot instantiation from context, and presents the offline meta-training objective.
Section~\ref{s:results} reports the numerical experiments, comparisons with the baseline models, and a discussion of the study limitations.
Finally, Section~\ref{s:conclusion} summarizes the main findings and outlines future research directions.

\paragraphfont{Notation}
We denote by $\R$, $\R_{\geq 0}$, $\R_{>0}$, $\N$, and $\Zpos$ the sets of real, nonnegative real, strictly positive real, natural numbers (including zero), and strictly positive integers, respectively.
For integers $a, b \in \N$ with $a \leq b$, the discrete index set is $\iset{a}{b} \coloneq \set*{a, a+1, \dots, b}{}{}$.
Vectors are typeset in boldface lowercase (e.g., $\bm{v} \in \R^n$, with $n \in \Zpos$), and matrices in uppercase (e.g., $W \in \R^{n \times m}$, with $n,m \in \Zpos$).
The operator $\norm{\cdot}$ denotes vector Euclidean norm, scalar absolute value, or set cardinality, and $\norm{A}_{\mathrm{F}} \coloneq \sqrt{\sum_{j,l} A_{jl}^2}$ is the Frobenius norm of $A$.
%
%
%
For discrete-time signals $\bm{s}_k$ sampled at step $k \in \N$, $\pc{\bm{s}_k}_{k=k_1}^{k_2}$ denotes the sequence over $\iset{k_1}{k_2}$, and $\bar{\bm{s}}$ denotes the temporal sample mean.
The function composition operator is $\circ$.
All-zero and all-one vectors of appropriate dimensions are denoted by $\bm{0}$ and $\bm{1}$, respectively.
%

\section{Problem Statement}
\label{s:ssae}
\label{s:problem_statement}
%
We consider the identification of a family of discrete-time nonlinear systems with input $\bm{u}_k^{(i)} \in \R^{n_u}$ and output $\bm{y}_k^{(i)} \in \R^{n_y}$, where $(i)$ indexes the system instance.
We assume that all members of the family share the state dimension $n_s \in \Zpos$ and the input and output dimensions $n_u,n_y \in \Zpos$, while their dynamics depend on a time-invariant parameter vector $\bm{\xi}^{(i)} \in \Xi \subset \R^{n_\xi}$, with $n_\xi \in \Zpos$:
\begin{equation*}
    \begin{cases}
        \bm{s}_{k+1}^{(i)} = h\pr*{\bm{s}_k^{(i)}, \bm{u}_k^{(i)}; \bm{\xi}^{(i)}}, \\
        \bm{y}_k^{(i)} = g\pr*{\bm{s}_k^{(i)}; \bm{\xi}^{(i)}} + \bm{v}_k^{(i)},
    \end{cases}
\end{equation*}
where $\bm{s}_k^{(i)} \in \R^{n_s}$ is the unobserved physical state and $\bm{v}_k^{(i)} \in \R^{n_y}$ is measurement noise.
We assume that the maps $h \colon \R^{n_s}\times\R^{n_u}\to\R^{n_s}$ and $g \colon \R^{n_s}\to\R^{n_y}$ are unknown and that the system parameters $\bm{\xi}^{(i)}$ are unavailable.

For each target system, we observe a finite input--output context trajectory of length $N_{\mathrm{c}} \in \Zpos$:
\begin{equation}
    \label{eq:dataset_context}
    \bcontext^{(i)} \coloneq \set*{\pr*{\bm{u}_k^{(i)},\bm{y}_k^{(i)}}}{k=0}{N_{\mathrm{c}}-1}.
\end{equation}
We aim to infer from $\bcontext^{(i)}$ the parameters of a compact state space model that we can directly use to predict and simulate independent trajectories of the same system.
We learn such context-to-parameter mapping offline, from data acquired on related systems, and apply it to previously unseen instances of the target family.

\section{Transfer-Meta State Space Autoencoders}
\label{s:meta_ssae}
%
\begin{figure*}[t]
    \centering
    \begin{tikzpicture}[
        every node/.style={font=\scriptsize},
        box/.style={draw, rounded corners=3pt, thick, align=center, inner xsep=2pt, inner ysep=1.5pt},
        frozen_box/.style={box, draw=cyan!70!blue, fill=cyan!8},
        train_box/.style={box, draw=orange!80!red, fill=orange!8},
        gray_box/.style={box, draw=gray!60, fill=gray!8},
        arrow/.style={->, >=stealth, thick, draw=gray!80!black, rounded corners=3pt},
        param_arrow/.style={->, >=stealth, dashed, draw=orange!85!red, line width=0.8pt, rounded corners=3pt},
        node distance=0.25cm and 0.8cm
    ]


    \coordinate (origin) at (0, 2.15);

    \draw[dashed, draw=blue!50!gray, fill=blue!2, rounded corners=5pt, line width=0.7pt] 
        (origin) rectangle ++(6.0, -5.4);
    \node[font=\footnotesize\bfseries, text=blue!70!black, anchor=north] 
        at ([xshift=3.0cm, yshift=-0.1cm]origin) {Context-to-Parameter Mapping};

    \coordinate (c2_left)  at ([xshift=6.8cm]origin);
    \coordinate (c2_right) at ([xshift=17.6cm]origin);
    
    \draw[dashed, draw=orange!60!gray, fill=orange!2, rounded corners=5pt, line width=0.7pt] 
        (c2_left) rectangle ([yshift=-5.4cm]c2_right);
    \node[font=\footnotesize\bfseries, text=red!70!black, anchor=north] 
        at ($([yshift=-0.1cm]c2_left)!0.5!([yshift=-0.1cm]c2_right)$) {Instantiated State Space Autoencoder $\Sigma\pr{\bparameters_{\bembedding^{(i)}}}$};

    \coordinate (mid_corr) at ($([xshift=6.0cm]origin)!0.5!(c2_left)$);


    \node[gray_box, minimum width=4.8cm, minimum height=0.65cm] (supp) at ([xshift=3.0cm, yshift=-1.0cm]origin) {
        $\bcontext^{(i)} \coloneq \set*{\pr*{\bm{u}_k^{(i)}, \bm{y}_k^{(i)}}}{k=0}{N_{\mathrm{c}}-1}$
    };

    \node[frozen_box, minimum width=4.8cm, minimum height=0.65cm, below=0.4cm of supp] (encoder) {
        \textcolor{cyan!70!blue}{\faSnowflake}\ \textbf{Frozen Pretrained Temporal Encoder} \eqref{eq:encoder_eval}\\[1pt]
        $H^{(i)} = \Gamma_{\mathrm{frozen}}\pr*{\bcontext^{(i)}}$
    };

    \node[train_box, minimum width=4.8cm, minimum height=0.65cm, below=0.4cm of encoder] (comp) {
        \textcolor{orange!90!red}{\faFire}\ \textbf{Context Compressor} \eqref{eq:context_compression}\\[1pt]
        $\bembedding^{(i)} = \Gamma_{\mathrm{comp}}\pr*{H^{(i)};\; \bparameters_{\Gamma_{\mathrm{comp}}}}$
    };

    \node[train_box, minimum width=4.8cm, minimum height=0.65cm, below=0.4cm of comp] (manifold) {
        \textcolor{orange!90!red}{\faFire}\ \textbf{Affine Parameter Maps} \eqref{eq:affine_maps}\\[1pt]
        $\bparameters_{m,\bembedding^{(i)}} = \bparameters_{m,0} + L_m \bembedding^{(i)}, \;\; \forall m \in \pc*{e, f, d}$
    };

    \draw[arrow] (supp) -- (encoder) node[pos=0.4, xshift=12pt, font=\tiny] {$\bcontext^{(i)}$};
    \draw[arrow] (encoder) -- (comp) node[pos=0.4, xshift=12pt, font=\tiny] {$H^{(i)}$};
    \draw[arrow] (comp) -- (manifold) node[pos=0.4, xshift=12pt, font=\tiny] {$\bembedding^{(i)}$};


    \node[gray_box, inner sep=3pt, minimum width=0.85cm] (io_in) at ([xshift=0.6cm, yshift=-1.45cm]c2_left) {$\binformation_{k-1}^{(i)}$};
    
    \node[train_box, minimum width=3.6cm, minimum height=0.72cm, right=0.35cm of io_in] (enc_map) {
        \textcolor{orange!90!red}{\faFire}\ \textbf{State Encoder} \eqref{eq:e}\\[1pt]
        $\bm{x}_k^{(i)} = e\pr*{\binformation_{k-1}^{(i)}; \bparameters_{e,\bembedding^{(i)}}}$
    };

    \node[train_box, minimum width=3.6cm, minimum height=0.72cm, below=0.75cm of enc_map] (trans_map) {
        \textcolor{orange!90!red}{\faFire}\ \textbf{State Transition Map} \eqref{eq:f}\\[1pt]
        $\tilde{\bm{x}}_{k+1}^{(i)} = f\pr*{\bm{x}_k^{(i)}, \bm{u}_k^{(i)}; \bparameters_{f,\bembedding^{(i)}}}$
    };

    \node[gray_box, inner sep=3pt, minimum width=0.85cm] (u_in) at (io_in |- trans_map) {$\bm{u}_k^{(i)}$};

    \node[train_box, minimum width=3.6cm, minimum height=0.72cm, right=0.7cm of $(enc_map.east)!0.5!(trans_map.east)$] (dec_map) {
        \textcolor{orange!90!red}{\faFire}\ \textbf{Output Decoder} \eqref{eq:d}\\[1pt]
        $\hat{\boutputvec}_k^{(i)} = d\pr*{\bm{x}_k^{(i)};\; \bparameters_{d,\bembedding^{(i)}}}$
    };

    \node[gray_box, inner sep=3pt, right=0.35cm of dec_map, align=center] (y_out) {
        {$\hat{\bm{y}}_k^{(i)}$}
    };

    \draw[arrow] (io_in) -- (enc_map);
    \draw[arrow] (u_in) -- (trans_map);

    \draw[arrow, dashed] ([xshift=-20pt]enc_map.south) -- ([xshift=-20pt]trans_map.north) node[midway, left=1pt, font=\tiny] {$\bm{x}_{k}^{(i)}$};

    \draw[arrow] (enc_map.east) -- ++(0.35cm, 0) node[midway, xshift=1.1cm, yshift=0.1cm, font=\tiny] {$\bm{x}_k^{(i)}$ (prediction)} |- ([yshift=6pt]dec_map.west);
    \draw[arrow, dashed] ([yshift=4pt]trans_map.east) -- ++(0.35cm, 0) |- ([yshift=-6pt]dec_map.west);

    \draw[arrow, dashed] ([yshift=-6pt]trans_map.east) -- ++(0.35, 0)  |- ([yshift=-8pt]trans_map.south east) node[midway, xshift=0.95cm, yshift=0.7cm, font=\tiny] {$\tilde{\bm{x}}_{k+1}^{(i)}$ (simulation)} -| ([xshift=20pt]trans_map.south);

    \draw[arrow] (dec_map) -- (y_out);


    \coordinate (bot_corr) at ([yshift=-22pt]trans_map.south);
    \coordinate (top_corr) at ([yshift=12pt]enc_map.north);

    \draw[param_arrow] (manifold.east) -- (manifold.east -| mid_corr) -- (top_corr -| mid_corr) -| (enc_map.north) node[pos=0.7, xshift= 18pt, font=\tiny, text=orange!90!red] {$\bparameters_{e,\bembedding^{(i)}}$};

    \draw[param_arrow] (manifold.east) -| (trans_map.south) node[pos=0.85, xshift=-18pt, font=\tiny, text=orange!90!red] {$\bparameters_{f,\bembedding^{(i)}}$};

    \draw[param_arrow] (manifold.east) -| (dec_map.south) node[pos=0.92, xshift=18pt, font=\tiny, text=orange!90!red] {$\bparameters_{d,\bembedding^{(i)}}$};

    \end{tikzpicture}
    \caption{Architecture of the proposed Transfer-Meta SSAE.
    The context-to-parameter mapping processes the measured context through a frozen pretrained temporal encoder and a context compressor to obtain a system embedding.
    Affine parameter maps generate the system-specific model parameters of the instantiated SSAE.
    Solid arrows denote the prediction pipeline at step $k$, where the state encoder reconstructs the latent state from measured history at each step.
    Dashed arrows denote the simulation loop, where the state is initialized by the encoder and recursively propagated through the state transition map under future inputs across successive steps.
    The output decoder reconstructs the output vector $\hat{\boutputvec}_k$, whose last $n_y$ entries give the estimated current output $\hat{\bm{y}}_k$.
    The snowflake (\textcolor{cyan!70!blue}{\smash{\faSnowflake}}) marks the frozen encoder; the flames (\textcolor{orange!90!red}{\smash{\faFire}}) mark components learned during target-family meta-training. 
    Model parameters remain fixed during target-system inference.
    }
    \label{fig:framework}
\end{figure*}
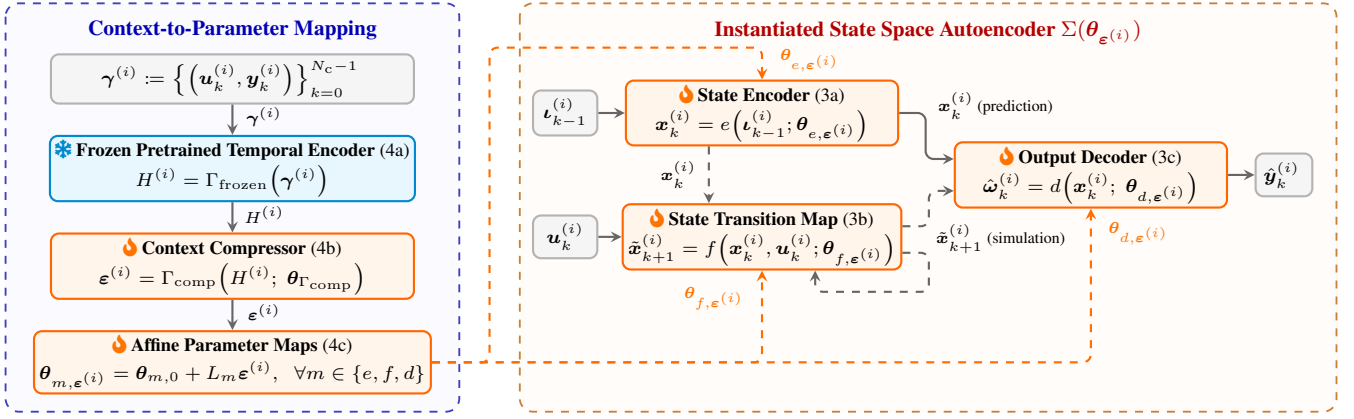

To obtain a zero-shot, in-context extension of the SSAE framework of~\citet{mastiLearningNonlinearState2021}, we learn a mapping from a context trajectory to the parameters of its state space maps.
First, we extract a compact characterization of the system from its measured context. 
Second, we use this characterization to determine the parameters of the SSAE model specialized for that system.
Finally, we initialize the SSAE latent state from a measured input--output history of the system and use the \emph{instantiated model} for prediction and simulation.
Here, we refer to an \emph{instantiated model} as the state space model whose parameters are specialized for a single target system instance.
We describe the state space maps, the context-to-parameter mapping, and the offline training objective below.

\subsection{State Space Model}
\label{ss:ssae_model}
%
From an architectural perspective, we retain the three maps of the SSAE architecture from~\citet{mastiLearningNonlinearState2021}: a state encoder $e$, a state transition map $f$, and an output decoder $d$.
Because the true state is unobserved, we learn a latent state $\bm{x}_k \in \R^{n_x}$, where the latent dimension $n_x \in \Zpos$ does not necessarily coincide with the physical state dimension $n_s$.
For a fixed system instance, we omit the superscript $(i)$ and define the measured input--output history vector and the output reconstruction vector as:
\begin{subequations}
    \begin{align}
        \binformation_k &\coloneq \ps*{ \bm{u}_{k-n_\text{a}+1}^\top, \dots, \bm{u}_k^\top, \; \bm{y}_{k-n_\text{b}+1}^\top, \dots, \bm{y}_k^\top }^\top \in \R^{n_{\information}}, \label{eq:info_vec}\\
        \boutputvec_k &\coloneq \ps*{ \bm{y}_{k-n_\text{c}+1}^\top, \dots, \bm{y}_k^\top }^\top \in \R^{n_{\outputvec}}, \label{eq:out_vec}
    \end{align}
\end{subequations}
where $n_\text{a},n_\text{b} \in \Zpos$ are the input and output history lengths,  $n_\text{c} \in \Zpos$ is the output reconstruction length, $1 \leq n_\text{c} \leq n_\text{b}$, $n_{\information}=n_\text{a}n_u+n_\text{b}n_y$, and $n_{\outputvec}=n_\text{c}n_y$.
We set the initial valid prediction index to $k_0 \coloneq \max\{n_\text{a},n_\text{b}\}$.
For a given model parameter set $\bparameters \coloneq \{\bparameters_e,\bparameters_f,\bparameters_d\}$, the state space model is therefore defined as:
\begin{subnumcases}{\Sigma\pr*{\bparameters}\coloneq\label{eq:ssae_equations}}
    \bm{x}_k = e\pr*{\binformation_{k-1}; \bparameters_e}, \label{eq:e}\\
    \tilde{\bm{x}}_{k+1} = f\pr*{\bm{x}_k, \bm{u}_k; \bparameters_f}, \label{eq:f}\\
    \hat{\boutputvec}_k = d\pr*{\bm{x}_k; \bparameters_d} \label{eq:d}.
\end{subnumcases}
We use $e \colon \R^{n_{\information}} \to \R^{n_x}$ to reconstruct the state from the measured history $\binformation_{k-1}$ and $f \colon \R^{n_x}\times\R^{n_u} \to \R^{n_x}$ to propagate it under the applied input.
The output decoder $d \colon \R^{n_x} \to \R^{n_{\outputvec}}$ provides the output reconstruction vector estimate $\hat{\boutputvec}_k$; the last $n_y$ components of $\hat{\boutputvec}_k$ provide the estimated current output $\hat{\bm{y}}_k$.
The remaining $(n_\text{c}-1)n_y$ past output components in $\hat{\boutputvec}_k$ help regularize the latent state $\bm{x}_k$ by forcing it to retain historical dynamics~\citep{mastiLearningNonlinearState2021}.
For a fixed system instance, the original SSAE framework learns $\bparameters$ through a dedicated optimization problem.
We instead infer these parameters directly from the context while retaining the same roles for $e$, $f$, and $d$.

\subsection{Transfer-Meta Parameterization}
\label{ss:meta_architecture}
%
To extend the SSAE scheme to a transfer meta-learning setting, we adopt the pipeline shown in \figurename{}~\ref{fig:framework}.
The pipeline comprises two functional blocks.
The context-to-parameter block, on the left, processes the context trajectory $\bcontext^{(i)}$ defined in~\eqref{eq:dataset_context} to obtain system-specific model parameters through a compact system embedding.
The SSAE block, on the right, uses these parameters in the state encoder $e$, state transition map $f$, and output decoder $d$ introduced in~\eqref{eq:ssae_equations}.
An affine parameter mapping connects the system embedding to the parameters of these three maps.
Thus, the context determines the parameters of the instantiated SSAE, while its maps retain their original roles and inputs.

The context-to-parameter mapping is:
\begin{subequations}
    \label{eq:meta_ssae_form}
    \begin{align}
        H^{(i)} &= \Gamma_{\mathrm{frozen}}\pr*{\bcontext^{(i)}}, \label{eq:encoder_eval}\\
        \bembedding^{(i)} &= \Gamma_{\mathrm{comp}}\pr*{H^{(i)};\; \bparameters_{\Gamma_{\mathrm{comp}}}} = \Gamma\pr*{\bcontext^{(i)};\; \bparameters_{\Gamma_{\mathrm{comp}}}}, \label{eq:context_compression}\\
        \bparameters_{m,\bembedding^{(i)}} &= \bparameters_{m,0}+L_m\bembedding^{(i)}, \quad m \in \{e,f,d\}. \label{eq:affine_maps}
    \end{align}
\end{subequations}
We first process the context through a frozen pretrained temporal encoder $\Gamma_{\mathrm{frozen}} \colon (\R^{n_u} \times \R^{n_y})^{N_{\mathrm{c}}} \to \R^{N_H \times d_H}$ (with $N_H, d_H \in \Zpos$) in~\eqref{eq:encoder_eval}.
In this work, to answer \textbf{RQ1}, we focus on a pretrained, \emph{frozen} encoder from a sequence model, specialized for dynamical prediction/identification on a \emph{source system class}, which differs from the \emph{target system family} under study. Here, in particular, the target family collects systems sharing a common dynamical structure, while an individual member is referred to as a \emph{target system instance}.
The encoder produces the temporal representation matrix $H^{(i)}\in\R^{N_H\times d_H}$, containing $N_H$ temporal representation vectors of dimension $d_H$.
The pretrained temporal encoder parameters remain fixed during both target-family meta-training and target-system inference.

Then, in~\eqref{eq:context_compression}, the trainable context compressor $\Gamma_{\mathrm{comp}} \colon \R^{N_H \times d_H}$ $\to \R^{n_{\embedding}}$, parameterized by $\bparameters_{\Gamma_{\mathrm{comp}}} \in \R^{p_{\mathrm{comp}}}$ with $p_{\mathrm{comp}} \in \Zpos$, aggregates these temporal representations into a system embedding $\bembedding^{(i)} \in \R^{n_{\embedding}}$, with $n_{\embedding} \in \Zpos$.
Since the pretrained representations were learned on the source system class, we train the compressor to produce a compact system embedding that captures the dynamical variability across instances of the target system family.
Together, the encoder and compressor define the context-to-embedding mapping $\Gamma \coloneq \Gamma_{\mathrm{comp}}\circ\Gamma_{\mathrm{frozen}} \colon (\R^{n_u} \times \R^{n_y})^{N_{\mathrm{c}}} \to \R^{n_{\embedding}}$.

To instantiate the SSAE model for a specific system $(i)$ from its embedding representation $\bembedding^{(i)}$ within a shared parameter space, we retain a common topology and parameter count $p_m \in \Zpos$ for each map $m \in \{e,f,d\}$.
Following~\citet{forgioneManifoldMetalearningReducedcomplexity2025}, we generate the system-specific model parameters via the affine parameter map in~\eqref{eq:affine_maps} for each map $m$.
Here, $\bparameters_{m,0} \in \R^{p_m}$ is a shared base parameter vector, and $L_m \in \R^{p_m\times n_{\embedding}}$ is a parameter variation matrix whose columns define directions of variation in the parameter space of map $m$.
The components of the embedding vector $\bembedding^{(i)}$ act as coordinates that weigh these directions to obtain the system-specific model parameters.
We jointly learn the context compressor, the shared base parameter vectors, and the parameter variation matrices during target-family meta-training.
Collecting the resulting parameters in~\eqref{eq:meta_ssae_form} as $\bparameters_{\bembedding^{(i)}} \coloneq \{\bparameters_{e,\bembedding^{(i)}}, \bparameters_{f,\bembedding^{(i)}}, \bparameters_{d,\bembedding^{(i)}}\}$, we instantiate the target model in~\eqref{eq:ssae_equations} as $\Sigma\pr*{\bparameters_{\bembedding^{(i)}}}$, without target-specific optimization.
The instantiated SSAE is used for prediction and simulation as described in Section~\ref{ss:zero_shot_use}.
The encoder $\Gamma_{\mathrm{frozen}}$ in~\eqref{eq:encoder_eval} and the compressor $\Gamma_{\mathrm{comp}}$ in~\eqref{eq:context_compression} adopted in our experiments are later specified in Section~\ref{ss:model_parameterization}.

\subsection{Zero-Shot Model Instantiation and Use}
\label{ss:zero_shot_use}
%
For an unseen system, we instantiate $\Sigma\pr*{\bparameters_{\bembedding^{(i)}}}$ in a zero-shot fashion by evaluating~\eqref{eq:meta_ssae_form} once on the target system context $\bcontext^{(i)}$ in~\eqref{eq:dataset_context}.
We then use the instantiated SSAE for prediction and simulation, with fixed parameters and without reevaluating the context mapping afterwards.

For open-loop simulation, we initialize the latent state $\bm{x}_{k_0}^{(i)}$ at the initial valid prediction index $k_0$ specified in Section~\ref{ss:ssae_model} through $e$ using the measured history $\binformation_{k_0-1}^{(i)}$, propagate the state through $f$ using future inputs, and reconstruct the outputs through $d$.
For one-step prediction, we reconstruct the state from the measured history at each step before evaluating the output decoder.
\begin{remark}
    By retaining the state reconstruction, transition, and output maps of the SSAE framework, we also obtain a model directly usable in state estimation and model-based control formulations, as discussed by~\citet{mastiLearningNonlinearState2021}.\footnote{Specifically, \citet{mastiLearningNonlinearState2021} demonstrated in practice the direct use of such identified state space maps for closed-loop model predictive control and state reconstruction, without requiring secondary model conversions.}    
\end{remark}

\subsection{Meta-Training Problem and Learning Procedure }
\label{ss:meta_training}
%
\paragraphfont{Training data and parameters}
We learn the meta-model using a population $\mathcal{I}_{\mathrm{train}}$ of related system instances.
For each system, we use its context $\bcontext^{(i)}$ in~\eqref{eq:dataset_context} to instantiate $\Sigma\pr*{\bparameters_{\bembedding^{(i)}}}$ through~\eqref{eq:meta_ssae_form} and evaluate the identification performance on an independent training query trajectory:
\begin{equation}
    \label{eq:dataset_query_tr}
    \ds_{\mathrm{q}}^{(i)} \coloneq \set*{\pr*{\bm{u}_k^{\mathrm{q},(i)},\bm{y}_k^{\mathrm{q},(i)}}}{k=0}{N_{\mathrm{q}}-1}, \quad i \in \mathcal{I}_{\mathrm{train}}.
\end{equation}
We take $N_{\mathrm{q}} \in \Zpos$ and keep context and query trajectories separate so that the inferred parameters are trained to generalize to independent trajectories of the same system.
We jointly learn the context compressor $\Gamma_{\mathrm{comp}}$ in~\eqref{eq:context_compression} and the affine parameter maps in~\eqref{eq:affine_maps} through the shared trainable parameter set $\bParameters \coloneq \pc{\bparameters_{\Gamma_{\mathrm{comp}}},\; \pc*{\bparameters_{m,0},L_m}_{m\in\{e,f,d\}}}$.
This shared trainable parameter set contains $p_{\mathrm{comp}}+(1+n_{\embedding})\sum_{m\in\{e,f,d\}}p_m$ scalar parameters; we keep the pretrained encoder $\Gamma_{\mathrm{frozen}}$ in~\eqref{eq:encoder_eval} fixed.

\paragraphfont{Dynamic loss}
We retain the three SSAE loss terms of~\citet{mastiLearningNonlinearState2021} and evaluate them over query windows of length $W \coloneq k_0+P+1$, where $P \in \Zpos$ is the simulation horizon and $N_{\mathrm{q}} \geq W$.
We denote the set of valid initial prediction indices by $\mathcal{K}\coloneq\iset{k_0}{N_{\mathrm{q}}-P-1}$.
For each system $i \in \mathcal{I}_{\mathrm{train}}$ with instantiated model parameters $\bparameters_{\bembedding^{(i)}}$ and each initial prediction index $k\in\mathcal{K}$, we construct $\binformation_{k+p-1}^{(i)}$ and $\boutputvec_{k+p}^{(i)}$ for $p \in \iset{0}{P}$ from its query trajectory in~\eqref{eq:dataset_query_tr} according to~\eqref{eq:info_vec} and~\eqref{eq:out_vec}.
We define the encoded latent states $\bm{x}_{k+p}^{(i)} \coloneq e\pr{\binformation_{k+p-1}^{(i)}; \bparameters_{e,\bembedding^{(i)}}}$ and the simulated latent states $\tilde{\bm{x}}_{k+p|k}^{(i)}$, initialized at $\tilde{\bm{x}}_{k|k}^{(i)} = \bm{x}_k^{(i)}$ and propagated through the state transition map, i.e., $\tilde{\bm{x}}_{k+p+1|k}^{(i)} = f\pr{\tilde{\bm{x}}_{k+p|k}^{(i)}, \bm{u}_{k+p}^{(i)}; \bparameters_{f,\bembedding^{(i)}}}$. 
The dynamic loss $\loss{\mathrm{dyn}}^{(i)}\pr*{k; \bparameters_{\bembedding^{(i)}}} \coloneq \sum_{\ell=1}^3 \lossw{\ell} \loss{\ell}^{(i)}\pr*{k; \bparameters_{\bembedding^{(i)}}}$ at the index $k \in \mathcal{K}$ for system $i \in \mathcal{I}_{\mathrm{train}}$ with weights $\lossw{\ell} \in \R_{\geq 0}$ for $\ell \in \iset{1}{3}$ combines the output reconstruction error $\loss{1}^{(i)}$, latent state consistency $\loss{2}^{(i)}$, and simulation error $\loss{3}^{(i)}$:
\begin{align*}
    \loss{1}^{(i)}\pr*{k; \bparameters_{\bembedding^{(i)}}} &\coloneq \frac{1}{(P+1) n_{\outputvec}} \sum_{p=0}^P \norm*{ d\pr*{\bm{x}_{k+p}^{(i)}; \bparameters_{d,\bembedding^{(i)}}} - \boutputvec_{k+p}^{(i)} }^2, \\
    \loss{2}^{(i)}\pr*{k; \bparameters_{\bembedding^{(i)}}} &\coloneq \frac{1}{P n_x} \sum_{p=1}^P \norm*{ \tilde{\bm{x}}_{k+p|k}^{(i)} - \bm{x}_{k+p}^{(i)} }^2, \\
    \loss{3}^{(i)}\pr*{k; \bparameters_{\bembedding^{(i)}}} &\coloneq \frac{1}{P n_{\outputvec}} \sum_{p=1}^P \norm*{ d\pr*{\tilde{\bm{x}}_{k+p|k}^{(i)}; \bparameters_{d,\bembedding^{(i)}}} - \boutputvec_{k+p}^{(i)} }^2.
\end{align*}

\paragraphfont{Embedding regularization}
We regularize the system embedding to remain consistent across different portions of the same context while retaining variability across systems.
For each system, we split the context $\bcontext^{(i)}$ in~\eqref{eq:dataset_context} into two non-overlapping temporal context views, $\bm\gamma_{\mathrm{a}}^{(i)}$ and $\bm\gamma_{\mathrm{b}}^{(i)}$, assuming the context length $N_{\mathrm{c}}$ is even:
\begin{equation*}
    \bcontext_{\mathrm{a}}^{(i)} \!\!\coloneq\!\! \set*{\!\pr*{\bm{u}_k^{(i)},\bm{y}_k^{(i)}}\!}{k=0}{N_{\mathrm{c}}/2-1}\!\!\!, \, \bcontext_{\mathrm{b}}^{(i)} \!\!\coloneq\!\! \set*{\!\pr*{\bm{u}_k^{(i)},\bm{y}_k^{(i)}}\!}{k=N_{\mathrm{c}}/2}{N_{\mathrm{c}}-1}.
\end{equation*}
We compute $\bembedding_v^{(i)} \coloneq \Gamma\pr*{\bcontext_v^{(i)}; \bparameters_{\Gamma_{\mathrm{comp}}}}$ for $v \in \{\mathrm{a},\mathrm{b}\}$, as defined in Section~\ref{ss:meta_architecture}.
For each scalar coordinate $\ell \in \iset{1}{n_{\embedding}}$ of the system embedding, we denote its value by $\embedding_{\ell,v}^{(i)}$ and compute its scalar training-population mean and regularized standard deviation:
\begin{align*}
    \bar{\embedding}_{\ell,v} &\coloneq \frac{1}{\abs{\mathcal{I}_{\mathrm{train}}}}\sum_{i\in\mathcal{I}_{\mathrm{train}}}\embedding_{\ell,v}^{(i)},\\
    \sigma_{\ell,v}\pr*{\mathcal{I}_{\mathrm{train}}} &\coloneq \sqrt{\frac{1}{\abs{\mathcal{I}_{\mathrm{train}}}} \sum_{i\in\mathcal{I}_{\mathrm{train}}} \pr*{\embedding_{\ell,v}^{(i)} - \bar{\embedding}_{\ell,v}}^2 + 10^{-6}}.
\end{align*}
We define the within-system embedding consistency loss $\loss{\mathrm{inv}}$ and the between-system embedding spread loss $\loss{\mathrm{var}}$ as:
\begin{align*}
    \loss{\mathrm{inv}}\pr*{\mathcal{I}_{\mathrm{train}}; \bparameters_{\Gamma_{\mathrm{comp}}}} &\coloneq \frac{1}{\abs{\mathcal{I}_{\mathrm{train}}} n_{\embedding}} \sum_{i\in\mathcal{I}_{\mathrm{train}}} \norm{\bembedding_{\mathrm{a}}^{(i)} - \bembedding_{\mathrm{b}}^{(i)}}^2, \\
    \loss{\mathrm{var}}\pr*{\mathcal{I}_{\mathrm{train}}; \bparameters_{\Gamma_{\mathrm{comp}}}} &\coloneq \frac{1}{2 n_{\embedding}} \sum_{v\in\pc{\mathrm{a},\mathrm{b}}} \sum_{\ell=1}^{n_{\embedding}} \max\pr*{0,1-\sigma_{\ell,v}\pr*{\mathcal{I}_{\mathrm{train}}}}.
\end{align*}
While $\loss{\mathrm{inv}}$ promotes consistency across context views of the same system, $\loss{\mathrm{var}}$ prevents embedding collapse across systems~\citep{bardes2022vicreg}; its unit target standard deviation fixes the embedding coordinate scale without loss of generality, letting $L_m$ in~\eqref{eq:affine_maps} absorb parameter variations.

\paragraphfont{Joint objective}
We regularize the shared trainable parameter set $\bm\Theta$ through:
\begin{equation*}
    R\pr*{\bParameters} \coloneq \norm{\bparameters_{\Gamma_{\mathrm{comp}}}}^2 + \!\! \sum_{m\in\{e,f,d\}} \!\! \pr*{\norm{\bparameters_{m,0}}^2+\norm{L_m}_{\mathrm{F}}^2}.
\end{equation*}
With $\lambda_{\mathrm{inv}},\lambda_{\mathrm{var}},\lambda_{\mathrm{reg}} \in \R_{\geq 0}$, we minimize:
\begin{equation}
    \label{eq:opt_problem_meta}
    \begin{aligned}
        \min_{\bParameters} \quad & \frac{1}{\abs{\mathcal{I}_{\mathrm{train}}}\,\abs{\mathcal{K}}} \sum_{i\in\mathcal{I}_{\mathrm{train}}} \sum_{k\in\mathcal{K}} \loss{\mathrm{dyn}}^{(i)}\pr*{k; \bparameters_{\bembedding^{(i)}}} + \lambda_{\mathrm{reg}}R\pr*{\bParameters}\\
        & + \lambda_{\mathrm{inv}}\loss{\mathrm{inv}}\pr*{\mathcal{I}_{\mathrm{train}}; \bparameters_{\Gamma_{\mathrm{comp}}}} + \lambda_{\mathrm{var}}\loss{\mathrm{var}}\pr*{\mathcal{I}_{\mathrm{train}}; \bparameters_{\Gamma_{\mathrm{comp}}}},\\
        %
        %
        \mathrm{s.t.} \quad & \eqref{eq:meta_ssae_form}, \forall i \in \mathcal{I}_{\mathrm{train}}.
    \end{aligned}
\end{equation}
We compute the full-context embedding $\bembedding^{(i)}$ and its system-specific model parameters once per system and reuse them across its query windows.
We propagate gradients through the compressor, the affine parameter maps, and the SSAE state simulations, while keeping the pretrained temporal encoder fixed.
We reserve separate system populations for validation and final evaluation, as detailed in Section~\ref{s:results}.
%
\section{Numerical Results}
\label{s:results}
%
In this section, we evaluate Transfer-Meta SSAE on two nonlinear benchmark families: Buck--Boost converters and Bouc--Wen oscillators.
Specifically, each benchmark family is considered independently to assess the cross-class transfer capability of the frozen temporal encoder (pretrained on synthetic Wiener--Hammerstein systems~\citep{rufoloEnhancedTransformerArchitecture2025}).
Target-family meta-training is conducted separately for each family.
The experiments assess zero-shot prediction and simulation performance, comparing the proposed approach with individually fitted baseline models on held-out system instances.
%
\subsection{Benchmark Families}
\label{ss:benchmarks}
%
\begin{table}[t]
    \centering
    \caption{Benchmark parameters: intervals are sampled uniformly, while scalar values are held fixed across the family.}
    \label{tab:benchmark_ranges}
    \settablestretch
    \scriptsize
    \setlength{\tabcolsep}{2.5pt}
    \begin{tabular*}{\columnwidth}{@{\extracolsep{\fill}} l l l c l l l @{}}
        \toprule
        \multicolumn{3}{c}{\textbf{Buck--Boost}} & & \multicolumn{3}{c}{\textbf{Bouc--Wen}} \\
        \cmidrule{1-3} \cmidrule{5-7}
        \textbf{Parameter} & \textbf{Value} & \textbf{Unit} & & \textbf{Parameter} & \textbf{Value} & \textbf{Unit} \\
        \midrule
        $L$                                & $\ps*{95,165}$        & $\mu\mathrm{H}$        & & $m_{\mathrm{L}}$          & $\ps*{1,3}$                & $\mathrm{kg}$ \\
        $C_{\mathrm{o}}$                   & $\ps*{11,19}$         & $\mathrm{mF}$           & & $c_{\mathrm{L}}$          & $\ps*{5,15}$               & $\mathrm{N\,s/m}$ \\
        $C_{\mathrm{i}}$                   & $\ps*{0.75,1.25}$     & $\mathrm{mF}$           & & $k_{\mathrm{L}}, \alpha$  & $\ps*{2.5,7.5}\times10^4$  & $\mathrm{N/m}$ \\
        $R_{\mathrm{L}}$                   & $\ps*{6,14}$          & $\mathrm{m}\Omega$      & & $\beta$                   & $\ps*{500,4500}$           & $\mathrm{m}^{-1}$ \\
        $R_{\mathrm{p}}$                   & $\ps*{0.35,0.75}$     & $\Omega$                & & $\gamma$                  & $\ps*{0.5,0.9}$            & $-$ \\
        $V_{\mathrm{p}}, I_{\mathrm{o}}$   & $200,\,80$            & $\mathrm{V},\,\mathrm{A}$ & & $\delta$                  & $\ps*{-1.5,-0.5}$          & $-$ \\
        $R_{\mathrm{Ci}}, R_{\mathrm{Co}}$ & $74,\,5$              & $\mathrm{m}\Omega$      & & $\nu$                     & $1$                        & $-$ \\
        \bottomrule
    \end{tabular*}
\end{table}
%
\subsubsection{Buck--Boost Converter}
%
The Buck--Boost benchmark is a non-isolated switching DC--DC power converter operating in continuous-conduction mode.
Its circuit topology and nominal operating specifications follow the Texas Instruments reference design of~\citet{bhardwaj2015modeling}.
The three physical states are the output capacitor voltage $v_{\mathrm{Co}}$ and input capacitor voltage $v_{\mathrm{Ci}}$, both in V, and the inductor current $i_{\mathrm{L}}$ in A.
The input $u_k$ is the transistor duty cycle and the measured output is $y_k=v_{\mathrm{Co},k}$.
The input is generated as a multisine signal comprising $100$ sinusoidal components with independently sampled phases at frequencies linearly spaced between $20$ Hz and $1\,500$ Hz.
The resulting signal is centred at $0.50$, scaled to have standard deviation $0.16$, and clipped to the interval $\ps*{0.10,0.90}$.
Each instance is parameterized by $n_\xi = 5$ parameters, namely the inductance $L$, capacitances $C_{\mathrm{o}}, C_{\mathrm{i}}$, and resistances $R_{\mathrm{L}}, R_{\mathrm{p}}$:
\begin{equation*}
    \bm{\xi}^{(i)} \coloneq \ps*{ L^{(i)}, \; C_{\mathrm{o}}^{(i)}, \; C_{\mathrm{i}}^{(i)}, \; R_{\mathrm{L}}^{(i)}, \; R_{\mathrm{p}}^{(i)} }^\trans \in \R_{>0}^5.
\end{equation*}
We sample these parameters independently and uniformly over the ranges reported in Table~\ref{tab:benchmark_ranges}.
The other circuit parameters are fixed to the values reported in the same table.
We integrate the continuous-time dynamics using fourth-order Runge--Kutta at $10$ kHz and discard an initial transient of $1\,000$ samples.
The continuous-time model is reported in Appendix~\ref{ss:app_buck_boost}.
%
\subsubsection{Bouc--Wen Oscillator}
%
The Bouc--Wen benchmark describes a vibrating mechanical oscillator with nonlinear hysteretic restoring forces.
The three physical states are the displacement $p$ in m, the velocity $v=\dot{p}$ in m/s, and the unobserved hysteretic force state $z$ in N.
The input $u_k$ is the applied force in N and the measured output is $y_k=p_k$.
The input is generated as a multisine force comprising $146$ sinusoidal components with independently sampled phases at integer frequencies from $5$ Hz to $150$ Hz, scaled to a root-mean-square amplitude of $50$ N.
Each instance is parameterized by $n_\xi = 8$ parameters collecting the mass $m_{\mathrm{L}}$, damping $c_{\mathrm{L}}$, stiffness $k_{\mathrm{L}}$, and hysteresis parameters $\alpha, \beta, \gamma, \delta, \nu$:
\begin{equation*}
    \bm{\xi}^{(i)} \coloneq \ps*{ m_{\mathrm{L}}^{(i)}, \; c_{\mathrm{L}}^{(i)}, \; k_{\mathrm{L}}^{(i)}, \; \alpha^{(i)}, \; \beta^{(i)}, \; \gamma^{(i)}, \; \delta^{(i)}, \; \nu^{(i)} }^\trans \in \R^8,
\end{equation*}
We sample seven of these eight parameters independently and uniformly over the ranges reported in Table~\ref{tab:benchmark_ranges}, while keeping the hysteresis exponent $\nu$ fixed.
We integrate the continuous-time dynamics using fourth-order Runge--Kutta at $750$ Hz after discarding an initial transient of $200$ samples.
The parameterization, excitation, and sampling frequency follow~\citet{forgioneManifoldMetalearningReducedcomplexity2025}.
The continuous-time model is reported in Appendix~\ref{ss:app_bouc_wen}.
%
\subsection{Population Split and Data Generation}
%
For each benchmark family, we independently sample mutually disjoint populations of $\abs{\mathcal{I}_{\mathrm{train}}}=8\,192$ training systems, $\abs{\mathcal{I}_{\mathrm{val}}}=1\,024$ validation systems, and $\abs{\mathcal{I}_{\mathrm{test}}}=1\,024$ test systems.
For each system instance, we generate a context trajectory and an independent training query of the form defined in~\eqref{eq:dataset_query_tr}, with $N_{\mathrm{c}}=N_{\mathrm{q}}=1\,024$.
The contexts and training queries of $\mathcal{I}_{\mathrm{train}}$ are used for meta-training.
For $i\in\mathcal{I}_{\mathrm{val}}$, the corresponding $\ds_{\mathrm{q}}^{(i)}$, together with its context, is used for validation-based model selection without parameter updates.
For each system instance, we also generate a test query with the same structure and length as the training query, independently of both the context and the training query.
Only the test queries of $\mathcal{I}_{\mathrm{test}}$ are used for final evaluation.
%
\subsubsection{Data Preprocessing}
%
For Buck--Boost, we express every output trajectory as a deviation from the equilibrium output $y_{\mathrm{eq}}^{(i)}$ at the nominal duty cycle $u_{\mathrm{nom}}=0.50$.
This preprocessing uses the benchmark parameters, which are not supplied to the context-to-parameter mapping.
Experimentally, the equilibrium output can instead be estimated from steady-state measurements at the nominal duty cycle.

For both benchmark families, we add zero-mean Gaussian measurement noise to the context and training query outputs.
For each system, the noise standard deviation is set to $5\%$ of the standard deviation of the corresponding noise-free context output, yielding an approximate signal-to-noise ratio of $26$~dB.
Test query outputs remain noise-free.

We then standardize the signals using statistics computed exclusively from the training queries of $\mathcal{I}_{\mathrm{train}}$ and applied unchanged to all populations.
For Buck--Boost, the input is standardized as $\tilde{u}_k=(u_k-0.50)/0.16$, while the output uses the mean and standard deviation pooled over the noisy, equilibrium-subtracted training query outputs.
For Bouc--Wen, both input and output use their respective pooled training query statistics.
%
\subsection{Model Configuration and Training}
\label{ss:model_parameterization}
%
We use the same Transfer-Meta SSAE architecture and hyperparameters for both benchmark families.
Both families are single-input single-output systems, with $n_u=n_y=1$.
We set the input and output history lengths to $n_\text{a}=n_\text{b}=20$, the output reconstruction length to $n_\text{c}=10$, the latent state dimension to $n_x=5$, and the system embedding dimension to $n_{\embedding}=8$.
We implement the SSAE maps $e$, $f$, and $d$ as multilayer perceptrons with two hidden layers of $32$ units each and hyperbolic-tangent activations, which are small enough to be considered for embedded use applications as well as being smooth, ensuring compatibility with sensitivity-based control and estimation techniques.
Each map has the same topology across system instances, while its system-specific model parameters vary through the affine parameter maps.
Each instantiated SSAE contains $5\,556$ parameters, including biases.

\paragraphfont{Frozen encoder and context compressor}
We reuse the temporal encoder of~\citet{rufoloEnhancedTransformerArchitecture2025}, pretrained on synthetic Wiener--Hammerstein systems, and discard its sequence-prediction decoder.
As anticipated, \emph{its weights always remain fixed during target-family meta-training and target-system inference}.
The context compressor consists of three small feedforward networks.
The first maps each temporal representation to a lower-dimensional vector, while the second assigns a scalar score to each resulting vector.
These scores are converted into positive weights that sum to one, and the resulting vectors are combined through a weighted average.
This aggregation implements attention pooling.
The third network maps this average to the $n_{\embedding}$-dimensional system embedding.
Each network has one hidden layer of 64 units with Gaussian error linear unit activation.
The encoder preprocessing and compressor implementation are provided in the replication package~\citep{decarliTransferMetaLearningState2026}.

\paragraphfont{Meta-training and model selection}
We perform meta-training separately for each benchmark family for $450$ epochs using the Adam optimizer to solve the regularized meta-training problem in~\eqref{eq:opt_problem_meta} with balanced mini-batches of systems and query windows.
All data-dependent loss terms are approximated on these mini-batches.
The simulation horizon is $P=256$, yielding local query windows of length $W=k_0+P+1=277$; gradients are computed using truncated backpropagation through time over this horizon.
After each epoch, we evaluate the joint meta-training objective on a mini-batch from the validation population.
After completing all $450$ epochs, we select the parameter configuration with the lowest validation objective.
The data-generation, training, and evaluation code, including the numerical training settings, is available in the replication package~\citep{decarliTransferMetaLearningState2026}.
%
\subsection{Baseline Methods}
We compare Transfer-Meta SSAE with two baseline models fitted independently for each test system using its noisy training query trajectory: an individual SSAE and a Nonlinear AutoRegressive model with eXogenous input (NARX).
The individual SSAE uses the same history lengths, output reconstruction length, latent state dimension, and map architectures as the instantiated Transfer-Meta SSAE.
Its training query is split chronologically into the first $80\%$ for fitting and the final $20\%$ as a baseline validation segment, distinct from the validation system population.
We select the model parameters that achieve the lowest open-loop simulation root mean squared error on this segment.

The NARX baseline uses $n_\text{a}=20$ past inputs and $n_\text{b}=20$ past outputs (matching the SSAE history lengths), a fixed random hyperbolic-tangent feature projection, and output weights fitted by ridge regression to the complete training query.
Its numerical settings are documented in the replication package~\citep{decarliTransferMetaLearningState2026}.

\subsection{Evaluation Protocol and Metrics}
%
For each test system, Transfer-Meta SSAE instantiates its model parameters from the context alone, without target-specific optimization.
All methods are evaluated on the same independent, noise-free test query, which is excluded from fitting and model selection.
The first $k_0=\max\{n_\text{a},n_\text{b}\}=20$ samples supply the initial history and are excluded from metric computation; the evaluated samples are indexed by $k\in\iset{k_0}{1\,023}$.
In one-step \emph{prediction}, both SSAE models reconstruct their latent states at each step from the test input--output history, while NARX uses the corresponding input--output regressor.
In open-loop \emph{simulation}, each model is initialized from the test query prefix and then evaluated using future inputs without subsequent reference-output feedback.
For NARX, estimated outputs are recursively supplied to its regressor.
For each system $(i)$ and operating regime, let $\bm{y}^{(i)}$ and $\hat{\bm{y}}^{(i)}$ collect the noise-free reference and estimated outputs over the evaluated samples, and let $\bar{\bm{y}}^{(i)}$ denote the temporal sample mean of $\bm{y}^{(i)}$ replicated over those samples.
For one-step prediction, the decoder evaluates the encoded state, $\hat{\boutputvec}_k = d(\bm{x}_k; \bparameters_{d,\bembedding^{(i)}})$, whereas for open-loop simulation it evaluates the propagated simulated state, $\hat{\boutputvec}_{k+p} = d(\tilde{\bm{x}}_{k+p\mid k}; \bparameters_{d,\bembedding^{(i)}})$; in both cases, the current estimated output $\hat{\bm{y}}_k$ corresponds to the last $n_y$ entries of $\hat{\boutputvec}_k$, as defined in~\eqref{eq:out_vec}.
We quantify accuracy using the FIT index:
\begin{equation*}
    \mathrm{FIT}^{(i)} \coloneq 100 \pr*{ 1 - \frac{\norm*{\bm{y}^{(i)} - \hat{\bm{y}}^{(i)}}}{\norm*{\bm{y}^{(i)} - \bar{\bm{y}}^{(i)}}} } \quad [\%].
\end{equation*}
%
%
\subsection{Performance Comparison}
\label{ss:results}
%
\begin{table}[t]
    \centering
    \caption{Aggregate FIT indices on test systems.}
    \label{tab:results}
    \settablestretch
    \small
    \setlength{\tabcolsep}{1.2pt}
    \begin{tabular*}{\columnwidth}{@{\extracolsep{\fill}} c l l c c @{}}
        \toprule
        & \textbf{Regime} & \textbf{Model} & \textbf{Mean FIT $\pm$ Std [\%]} & \textbf{Median FIT [\%]} \\
        \midrule
        \multirow{6}{*}{\rotatebox{90}{\textbf{Buck--Boost}}}
        & \multirow{3}{*}{Simulation}
        & \textbf{Transfer-Meta SSAE}   & $\mathbf{83.1 \pm 6.1}$             & $\mathbf{84.4}$ \\
        & & Individual SSAE             & $-7.7 \pm 14.4$                     & $-2.7$ \\
        & & NARX                        & $28.1 \pm 14.5^\dagger$             & $27.7$ \\
        \cmidrule{2-5}
        & \multirow{3}{*}{Prediction}
        & Transfer-Meta SSAE            & $90.6 \pm 2.8$                      & $90.7$ \\
        & & Individual SSAE             & $13.0 \pm 28.6$                     & $10.2$ \\
        & & \textbf{NARX}               & $\mathbf{93.4 \pm 1.3}$             & $\mathbf{93.6}$ \\
        \midrule
        \multirow{6}{*}{\rotatebox{90}{\textbf{Bouc--Wen}}}
        & \multirow{3}{*}{Simulation}
        & \textbf{Transfer-Meta SSAE}   & $\mathbf{85.6 \pm 8.9}$             & $\mathbf{88.5}$ \\
        & & Individual SSAE             & $45.1 \pm 11.8$                     & $47.6$ \\
        & & NARX                        & $73.1 \pm 16.8$                     & $78.7$ \\
        \cmidrule{2-5}
        & \multirow{3}{*}{Prediction}
        & \textbf{Transfer-Meta SSAE}   & $\mathbf{95.9 \pm 2.2}$             & $\mathbf{96.5}$ \\
        & & Individual SSAE             & $61.4 \pm 9.4$                      & $62.6$ \\
        & & NARX                        & $94.4 \pm 2.6$                      & $95.2$ \\
        \bottomrule
    \end{tabular*}
    \vspace{2pt}
    \raggedright
    \footnotesize{$^\dagger$ For NARX on Buck--Boost, $191$ of $1\,024$ simulations ($18.7\%$) were numerically divergent (minimum raw $\text{FIT} = -3.4 \times 10^6$); statistics are computed over the $833$ remaining instances to avoid metric distortion.}
\end{table}

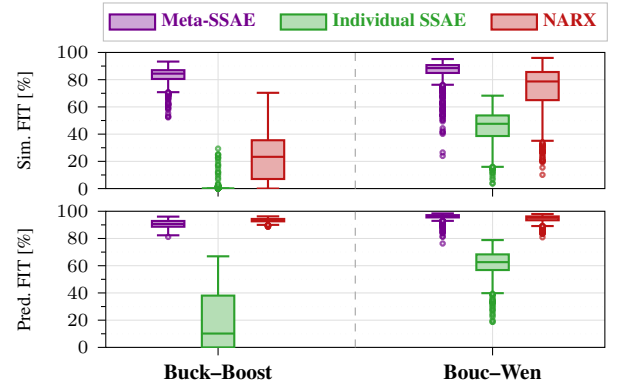
\begin{figure}[t]
    \centering
    \begin{tikzpicture}
        \pgfplotsset{
            boxplot style/.style={
                boxplot={
                    box extend=0.26,
                    whisker extend=0.18,
                    every median/.style={thick},
                    draw direction=y,
                },
                area legend,
                thick,
                mark=*,
                mark size=0.8pt,
                mark options={fill opacity=0.35, draw opacity=0.7},
            }
        }
        \begin{groupplot}[
            group style={
                group size=1 by 2,
                vertical sep=0.3cm,
                xlabels at=edge bottom,
                xticklabels at=edge bottom,
            },
            width=6.6cm,
            scale only axis,
            grid=both,
            ticks=both,
            tick align=outside,
            grid style={line width=0.2pt, draw=gray!25},
            minor grid style={line width=0.1pt, draw=gray!15, dotted},
            tick label style={font=\scriptsize},
            label style={font=\footnotesize},
            xmin=0.4, xmax=4.4,
            xtick={1.3, 3.5},
            enlarge x limits=false,
            ymin=0, ymax=100,
            enlarge y limits=false,
            ytick={0, 20, 40, 60, 80, 100},
            minor y tick num=1,   
            tick pos=left,
            major tick length=3.0pt,
            minor tick length=1.8pt,
            every minor tick/.style={draw=black, line width=0.4pt},
            every tick/.style={draw=black, line width=0.4pt},
        ]

            \nextgroupplot[
                height=1.8cm,
                minor y tick num=1,
                ylabel={\scriptsize Sim. FIT [\%]},
                ylabel shift=2pt,
                legend style={
                    at={(0.5, 1.08)},
                    anchor=south,
                    legend columns=3,
                    font=\scriptsize,
                    draw=gray!50,
                    fill=white,
                    inner sep=2pt,
                    /tikz/every even column/.append style={column sep=0.25cm},
                },
            ]
                \draw[dashed, gray!60, line width=0.5pt] (axis cs:2.4, 0) -- (axis cs:2.4, 105);

                \addplot+ [boxplot style, fill=meta_color!35, draw=meta_color, boxplot/draw position=0.9]
                    table [y=bbMetaSim] {data/boxplots.dat};
                \addlegendentry{\textcolor{meta_color}{\textbf{Meta-SSAE}}}

                \addplot+ [boxplot style, fill=indiv_color!35, draw=indiv_color, boxplot/draw position=1.3]
                    table [y=bbIndivSim] {data/boxplots.dat};
                \addlegendentry{\textcolor{indiv_color}{\textbf{Individual SSAE}}}

                \addplot+ [boxplot style, fill=narx_color!35, draw=narx_color, boxplot/draw position=1.7]
                    table [y=bbNarxSim] {data/boxplots.dat};
                \addlegendentry{\textcolor{narx_color}{\textbf{NARX}}}

                \addplot+ [boxplot style, fill=meta_color!35, draw=meta_color, boxplot/draw position=3.1, forget plot]
                    table [y=bwMetaSim] {data/boxplots.dat};
                \addplot+ [boxplot style, fill=indiv_color!35, draw=indiv_color, boxplot/draw position=3.5, forget plot]
                    table [y=bwIndivSim] {data/boxplots.dat};
                \addplot+ [boxplot style, fill=narx_color!35, draw=narx_color, boxplot/draw position=3.9, forget plot]
                    table [y=bwNarxSim] {data/boxplots.dat};


            \nextgroupplot[
                height=1.8cm,
                minor y tick num=1,
                ylabel={\scriptsize Pred. FIT [\%]},
                ylabel shift=2pt,
                xticklabels={\textbf{Buck--Boost}, \textbf{Bouc--Wen}},
                xticklabel style={font=\footnotesize},
            ]
                \draw[dashed, gray!60, line width=0.5pt] (axis cs:2.4, 0) -- (axis cs:2.4, 105);

                \addplot+ [boxplot style, fill=meta_color!35, draw=meta_color, boxplot/draw position=0.9, forget plot]
                    table [y=bbMetaStep] {data/boxplots.dat};
                \addplot+ [boxplot style, fill=indiv_color!35, draw=indiv_color, boxplot/draw position=1.3, forget plot]
                    table [y=bbIndivStep] {data/boxplots.dat};
                \addplot+ [boxplot style, fill=narx_color!35, draw=narx_color, boxplot/draw position=1.7, forget plot]
                    table [y=bbNarxStep] {data/boxplots.dat};

                \addplot+ [boxplot style, fill=meta_color!35, draw=meta_color, boxplot/draw position=3.1, forget plot]
                    table [y=bwMetaStep] {data/boxplots.dat};
                \addplot+ [boxplot style, fill=indiv_color!35, draw=indiv_color, boxplot/draw position=3.5, forget plot]
                    table [y=bwIndivStep] {data/boxplots.dat};
                \addplot+ [boxplot style, fill=narx_color!35, draw=narx_color, boxplot/draw position=3.9, forget plot]
                    table [y=bwNarxStep] {data/boxplots.dat};

        \end{groupplot}
    \end{tikzpicture}
    \vspace{-0.4cm}
    \caption{Distributions of the FIT index across held-out test systems for the two benchmark families.
    The top panel reports simulation FIT indices, whereas the bottom panel reports prediction FIT indices.}
    \label{fig:boxplots}
\end{figure}

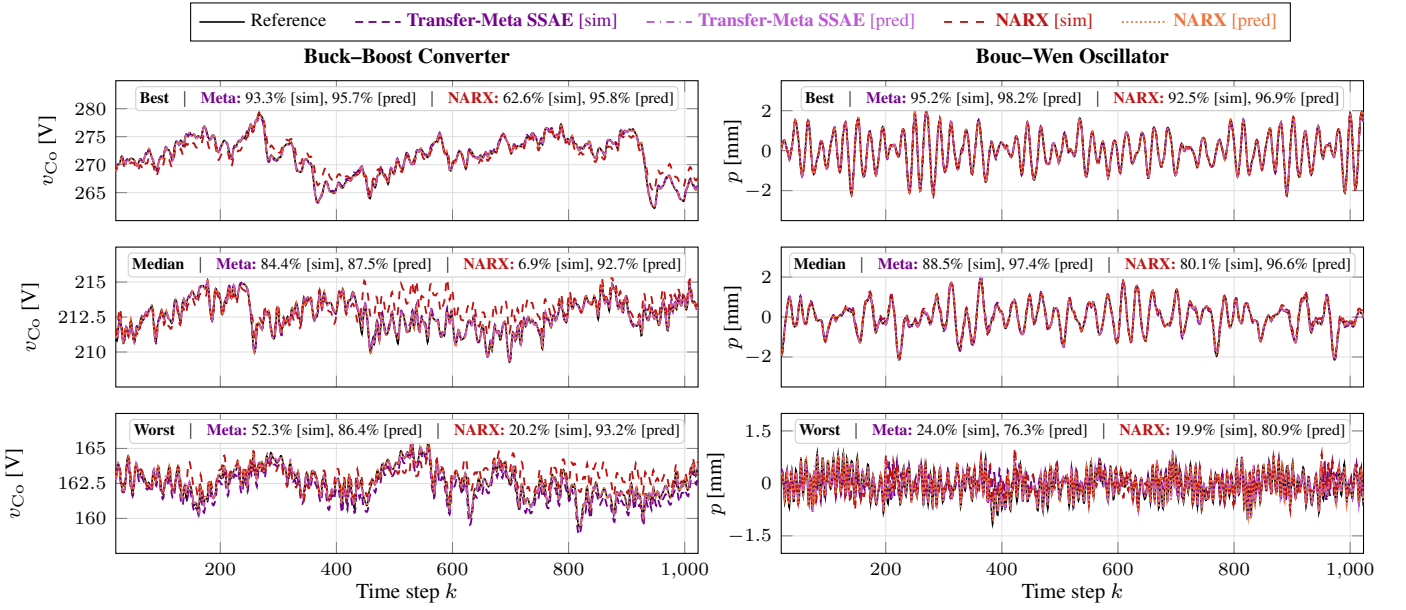
\begin{figure*}[ht]
    \centering
    \begin{tikzpicture}
        \begin{groupplot}[
            group style={
                group size=2 by 3,
                horizontal sep=1.1cm,
                vertical sep=0.35cm,
            },
            width=7.7cm,
            height=1.85cm,
            grid=both,
            grid style={line width=0.2pt, draw=gray!25},
            label style={font=\footnotesize},
            tick label style={font=\scriptsize},
            title style={font=\footnotesize\bfseries, yshift=-3pt},
            scale only axis,
            xmin=20, xmax=1023,
            xtick={200, 400, 600, 800, 1000},
        ]

            \nextgroupplot[
                title={\textbf{Buck--Boost Converter}},
                ylabel={$v_{\mathrm{Co}}$ [V]},
                ymin=260, ymax=285,
                ytick={265, 270, 275, 280},
                xticklabels={},
            ]
            \addplot[solid, color=real_color, line width=0.6pt] table [x=k, y=bbTrueB] {data/rollouts.dat};
            \addplot[densely dashed, color=meta_sim_color, line width=0.7pt] table [x=k, y=bbMetaSimB] {data/rollouts.dat};
            \addplot[dashdotted, color=meta_step_color, line width=0.7pt] table [x=k, y=bbMetaStepB] {data/rollouts.dat};
            \addplot[dashed, color=narx_sim_color, line width=0.7pt] table [x=k, y=bbNarxSimB] {data/rollouts.dat};
            \addplot[densely dotted, color=narx_step_color, line width=0.7pt] table [x=k, y=bbNarxStepB] {data/rollouts.dat};
            \node[anchor=north, font=\tiny, fill=white, fill opacity=0.92, text opacity=1, draw=gray!40, line width=0.3pt, rounded corners=1.5pt, inner sep=1.5pt] at (rel axis cs:0.50, 0.96) {
                \textbf{Best} \;\;$\mid$\;\;
                \textcolor{meta_sim_color}{\textbf{Meta:}} 93.3\% [sim], 95.7\% [pred] \;\;$\mid$\;\;
                \textcolor{narx_sim_color}{\textbf{NARX:}} 62.6\% [sim], 95.8\% [pred]
            };

            \nextgroupplot[
                title={\textbf{Bouc--Wen Oscillator}},
                ylabel style={yshift=-6pt},
                ylabel={$p$ [mm]},
                ymin=-3.5, ymax=3.5,
                ytick={-2, 0, 2},
                xticklabels={},
                legend style={
                    at={(-0.07, 1.28)},
                    anchor=south,
                    legend columns=5,
                    font=\scriptsize,
                    /tikz/every even column/.append style={column sep=0.30cm},
                },
            ]
            \addplot[solid, color=real_color, line width=0.6pt] table [x=k, y=bwTrueB] {data/rollouts.dat};
            \addlegendentry{Reference}
            \addplot[densely dashed, color=meta_sim_color, line width=0.7pt] table [x=k, y=bwMetaSimB] {data/rollouts.dat};
            \addlegendentry{\textcolor{meta_sim_color}{\textbf{Transfer-Meta SSAE} [sim]}}
            \addplot[dashdotted, color=meta_step_color, line width=0.7pt] table [x=k, y=bwMetaStepB] {data/rollouts.dat};
            \addlegendentry{\textcolor{meta_step_color}{\textbf{Transfer-Meta SSAE} [pred]}}
            \addplot[dashed, color=narx_sim_color, line width=0.7pt] table [x=k, y=bwNarxSimB] {data/rollouts.dat};
            \addlegendentry{\textcolor{narx_sim_color}{\textbf{NARX} [sim]}}
            \addplot[densely dotted, color=narx_step_color, line width=0.7pt] table [x=k, y=bwNarxStepB] {data/rollouts.dat};
            \addlegendentry{\textcolor{narx_step_color}{\textbf{NARX} [pred]}}
            \node[anchor=north, font=\tiny, fill=white, fill opacity=0.92, text opacity=1, draw=gray!40, line width=0.3pt, rounded corners=1.5pt, inner sep=1.5pt] at (rel axis cs:0.50, 0.96) {
                \textbf{Best} \;\;$\mid$\;\;
                \textcolor{meta_sim_color}{\textbf{Meta:}} 95.2\% [sim], 98.2\% [pred] \;\;$\mid$\;\;
                \textcolor{narx_sim_color}{\textbf{NARX:}} 92.5\% [sim], 96.9\% [pred]
            };

            \nextgroupplot[
                ylabel={$v_{\mathrm{Co}}$ [V]},
                ymin=207.5, ymax=217.5,
                ytick={210, 212.5, 215},
                xticklabels={},
            ]
            \addplot[solid, color=real_color, line width=0.6pt] table [x=k, y=bbTrueM] {data/rollouts.dat};
            \addplot[densely dashed, color=meta_sim_color, line width=0.7pt] table [x=k, y=bbMetaSimM] {data/rollouts.dat};
            \addplot[dashdotted, color=meta_step_color, line width=0.7pt] table [x=k, y=bbMetaStepM] {data/rollouts.dat};
            \addplot[dashed, color=narx_sim_color, line width=0.7pt] table [x=k, y=bbNarxSimM] {data/rollouts.dat};
            \addplot[densely dotted, color=narx_step_color, line width=0.7pt] table [x=k, y=bbNarxStepM] {data/rollouts.dat};
            \node[anchor=north, font=\tiny, fill=white, fill opacity=0.92, text opacity=1, draw=gray!40, line width=0.3pt, rounded corners=1.5pt, inner sep=1.5pt] at (rel axis cs:0.50, 0.96) {
                \textbf{Median} \;\;$\mid$\;\;
                \textcolor{meta_sim_color}{\textbf{Meta:}} 84.4\% [sim], 87.5\% [pred] \;\;$\mid$\;\;
                \textcolor{narx_sim_color}{\textbf{NARX:}} 6.9\% [sim], 92.7\% [pred]
            };

            \nextgroupplot[
                ylabel style={yshift=-6pt},
                ylabel={$p$ [mm]},
                ymin=-3.5, ymax=3.5,
                ytick={-2, 0, 2},
                xticklabels={},
            ]
            \addplot[solid, color=real_color, line width=0.6pt] table [x=k, y=bwTrueM] {data/rollouts.dat};
            \addplot[densely dashed, color=meta_sim_color, line width=0.7pt] table [x=k, y=bwMetaSimM] {data/rollouts.dat};
            \addplot[dashdotted, color=meta_step_color, line width=0.7pt] table [x=k, y=bwMetaStepM] {data/rollouts.dat};
            \addplot[dashed, color=narx_sim_color, line width=0.7pt] table [x=k, y=bwNarxSimM] {data/rollouts.dat};
            \addplot[densely dotted, color=narx_step_color, line width=0.7pt] table [x=k, y=bwNarxStepM] {data/rollouts.dat};
            \node[anchor=north, font=\tiny, fill=white, fill opacity=0.92, text opacity=1, draw=gray!40, line width=0.3pt, rounded corners=1.5pt, inner sep=1.5pt] at (rel axis cs:0.50, 0.96) {
                \textbf{Median} \;\;$\mid$\;\;
                \textcolor{meta_sim_color}{\textbf{Meta:}} 88.5\% [sim], 97.4\% [pred] \;\;$\mid$\;\;
                \textcolor{narx_sim_color}{\textbf{NARX:}} 80.1\% [sim], 96.6\% [pred]
            };

            \nextgroupplot[
                ylabel style={yshift=6pt},
                xlabel style={yshift=5pt},
                xlabel={Time step $k$},
                ylabel={$v_{\mathrm{Co}}$ [V]},
                ymin=157.5, ymax=167.5,
                ytick={160, 162.5, 165},
            ]
            \addplot[solid, color=real_color, line width=0.6pt] table [x=k, y=bbTrueW] {data/rollouts.dat};
            \addplot[densely dashed, color=meta_sim_color, line width=0.7pt] table [x=k, y=bbMetaSimW] {data/rollouts.dat};
            \addplot[dashdotted, color=meta_step_color, line width=0.7pt] table [x=k, y=bbMetaStepW] {data/rollouts.dat};
            \addplot[dashed, color=narx_sim_color, line width=0.7pt] table [x=k, y=bbNarxSimW] {data/rollouts.dat};
            \addplot[densely dotted, color=narx_step_color, line width=0.7pt] table [x=k, y=bbNarxStepW] {data/rollouts.dat};
            \node[anchor=north, font=\tiny, fill=white, fill opacity=0.92, text opacity=1, draw=gray!40, line width=0.3pt, rounded corners=1.5pt, inner sep=1.5pt] at (rel axis cs:0.50, 0.96) {
                \textbf{Worst} \;\;$\mid$\;\;
                \textcolor{meta_sim_color}{\textbf{Meta:}} 52.3\% [sim], 86.4\% [pred] \;\;$\mid$\;\;
                \textcolor{narx_sim_color}{\textbf{NARX:}} 20.2\% [sim], 93.2\% [pred]
            };

            \nextgroupplot[
                ylabel style={yshift=-6pt},
                xlabel style={yshift=5pt},
                xlabel={Time step $k$},
                ylabel={$p$ [mm]},
                ymin=-2.0, ymax=2.0,
                ytick={-1.5, 0, 1.5},
            ]
            \addplot[solid, color=real_color, line width=0.6pt] table [x=k, y=bwTrueW] {data/rollouts.dat};
            \addplot[densely dashed, color=meta_sim_color, line width=0.7pt] table [x=k, y=bwMetaSimW] {data/rollouts.dat};
            \addplot[dashdotted, color=meta_step_color, line width=0.7pt] table [x=k, y=bwMetaStepW] {data/rollouts.dat};
            \addplot[dashed, color=narx_sim_color, line width=0.7pt] table [x=k, y=bwNarxSimW] {data/rollouts.dat};
            \addplot[densely dotted, color=narx_step_color, line width=0.7pt] table [x=k, y=bwNarxStepW] {data/rollouts.dat};
            \node[anchor=north, font=\tiny, fill=white, fill opacity=0.92, text opacity=1, draw=gray!40, line width=0.3pt, rounded corners=1.5pt, inner sep=1.5pt] at (rel axis cs:0.50, 0.96) {
                \textbf{Worst} \;\;$\mid$\;\;
                \textcolor{meta_sim_color}{\textbf{Meta:}} 24.0\% [sim], 76.3\% [pred] \;\;$\mid$\;\;
                \textcolor{narx_sim_color}{\textbf{NARX:}} 19.9\% [sim], 80.9\% [pred]
            };

        \end{groupplot}
    \end{tikzpicture}
    \vspace{-0.60cm}
    \caption{Representative Buck--Boost and Bouc--Wen output trajectories for the held-out test systems with the best, median, and worst simulation FIT indices.
    Solid lines show the noise-free reference outputs; dashed and dotted lines show simulations and predictions, respectively, for Transfer-Meta SSAE and NARX.
    }
    \label{fig:rollouts}
\end{figure*}

Table~\ref{tab:results} reports the FIT statistics over the held-out test system instances, while \figurename{}~\ref{fig:boxplots} shows the corresponding FIT distributions.
Transfer-Meta SSAE achieves the highest median simulation FIT on both benchmark families: $84.4\%$ for Buck--Boost and $88.5\%$ for Bouc--Wen, compared with $27.7\%$ and $78.7\%$ for NARX and $-2.7\%$ and $47.6\%$ for individual SSAE, respectively.
Its simulation FIT distributions also remain concentrated at high values across both test populations.
NARX achieves prediction FIT comparable to Transfer-Meta SSAE, but, as expected for models fitted under the one-step-ahead prediction error criterion, its simulation FIT decreases substantially when estimated outputs are recursively fed back.
For Buck--Boost, $191$ NARX simulations are numerically divergent, so the simulation statistics in Table~\ref{tab:results} include only the remaining $833$ finite output trajectories.
Despite sharing the same SSAE map architectures, individual SSAE achieves substantially lower simulation FIT.
This is likely caused by the limited size of the individual training dataset, which may be insufficient to learn an accurate individual SSAE model independently.
Transfer-Meta SSAE appears to mitigate this data scarcity through a prior learned across the target system family, enabling zero-shot model instantiation from context.

The representative output trajectories in \figurename{}~\ref{fig:rollouts}, selected at the best, median, and worst simulation-performance percentiles, illustrate both the aggregate improvement and its limits.
For the median Buck--Boost case, Transfer-Meta SSAE reproduces the output dynamics during simulation, whereas the NARX output trajectory drifts despite its high prediction FIT.
For Bouc--Wen, Transfer-Meta SSAE reproduces the main amplitude and timing features across the representative cases.
Its worst-case simulation FIT nevertheless falls to $52.3\%$ for Buck--Boost and $24.0\%$ for Bouc--Wen, showing that zero-shot model instantiation does not eliminate difficult system instances.
Across the evaluated methods, the lowest simulation FIT values tended to occur for highly damped Buck--Boost converters with large $R_{\mathrm{p}}$ and for Bouc--Wen oscillators combining strong hysteretic dissipation, characterized by large $\beta/\lvert\delta\rvert$, with low mass $m_{\mathrm{L}}$.

Beyond the comparison with the baselines, the results support cross-class transfer of the frozen pretrained temporal encoder in the considered setting.
Although pretrained on Wiener--Hammerstein systems, its temporal representations remain useful for learning the context compressor and affine parameter maps separately for the Buck--Boost and Bouc--Wen target families.
%
\subsection{Computational Considerations}
\label{ss:computation}
%
Beyond transferring temporal representations, the proposed approach produces an explicit, compact system-specific model.
The system embedding determines the parameters of the instantiated SSAE, whose state space maps then perform prediction and simulation with fixed system-specific model parameters.
Crucially, the instantiated models $\Sigma(\bparameters_{\bembedding^{(i)}})$ contain only $5\,556$ parameters, including biases.
For perspective, the in-context sequence models evaluated on Wiener--Hammerstein systems by~\citet{rufoloEnhancedTransformerArchitecture2025} and~\citet{forgioneSystemModelsClass2023} contain $5.54$ million and approximately $5.6$ million parameters, respectively, for the simulation task, \emph{about three orders of magnitude more than the instantiated SSAE.}
Moreover, while the shared meta-parameter set $\bParameters$ contains $65\,149$ parameters (including the context compressor and the affine variation matrices), we stress that the context-to-parameter mapping is evaluated once at instantiation and never during runtime.
Note that this parameter count comparison highlights runtime model compactness and memory footprint, rather than benchmarking identification performance across different datasets.

The computational separation between one-time offline context processing (model parameter instantiation) and subsequent online model evaluation also determines when computational cost is incurred.
Target-family meta-training took approximately $45$~min per benchmark family, including the training and validation passes.\footnote{We ran the experiments on a machine equipped with an AMD Ryzen 9 7900X 12-core CPU, 32~GB of RAM, and an NVIDIA GeForce RTX 4070 Ti SUPER GPU with 16~GB of memory.
The meta-training code uses JAX/Equinox with the CUDA backend.}
In comparison, training reference in-context sequence predictors from scratch is typically a matter of days (e.g., $1.0$--$3.8$~days in~\citet{rufoloEnhancedTransformerArchitecture2025} and over $5$~days in~\citet{forgioneSystemModelsClass2023}).
Moreover, while a population of $8\,192$ systems may appear substantial, reference in-context sequence models typically rely on pretraining campaigns over tens of millions of synthetic systems (e.g., $32$~million generated system instances in~\citet{rufoloEnhancedTransformerArchitecture2025}).
Reusing the frozen source representation bypasses this burden for the target family.
\begin{remark}
    We stress that training a multi-million-parameter sequence predictor from scratch on only $8\,192$ target instances would be constrained by data scarcity, and, even if competitive, it would still not provide a compact state space model for simulation.
\end{remark}
\section{Conclusion}
\label{s:conclusion}
%
We presented Transfer-Meta SSAEs, which combine a frozen pretrained temporal encoder, a context compressor, and affine parameter maps to instantiate nonlinear state space models from input--output context trajectories.

The results indicate that a pretrained temporal encoder can extract useful temporal representations when transferred across distinct system classes without parameter updates, therefore answering affirmatively to \textbf{RQ1}.
%

%
The resulting models achieve higher median simulation FIT than individually fitted baselines on both target families while exposing an explicit state space model, therefore answering the model-instantiation question \textbf{RQ2}.

%
Future work will assess alternative temporal encoders and context compressor architectures, ablate the affine parameterization and context length, extend the approach to heterogeneous system families, and evaluate the instantiated models in state estimation and model predictive control applications.

\appendix
\section{Benchmark Models}
\label{s:appendix}
\begin{figure}[!t]
    \centering
    \resizebox{\columnwidth}{!}{%
    \begin{tikzpicture}[every node/.style={font=\scriptsize}]

        \begin{scope}[shift={(0,0)}]
            \node[font=\footnotesize\bfseries, anchor=center]
                at (4.1,2.5)
                {(a) Buck--Boost converter};

            \draw[thick] (0.2,1.8) -- (0.8,1.8);
            \filldraw[fill=white] (0.2,1.8) circle (1.5pt)
                node[left=1pt] {$V_{\mathrm{p}}$};

            \draw[thick, fill=white]
                (0.8,1.65) rectangle (1.4,1.95)
                node[pos=0.5, above=0.15cm] {$R_{\mathrm{p}}$};
            \draw[thick] (1.4,1.8) -- (2.2,1.8);

            \draw[thick] (2.2,1.8) -- (2.2,1.4);
            \draw[thick, fill=white]
                (1.95,1.1) rectangle (2.45,1.4)
                node[pos=0.5, right=0.3cm] {$R_{\mathrm{Ci}}$};
            \draw[thick] (2.2,1.1) -- (2.2,0.8);
            \draw[thick] (1.9,0.8) -- (2.5,0.8);
            \draw[thick] (1.9,0.65) -- (2.5,0.65)
                node[right=0.0cm] {$C_{\mathrm{i}}$};
            \draw[thick] (2.2,0.65) -- (2.2,0);

            \draw[thick] (2.2,1.8) -- (2.8,1.8);
            \draw[thick, fill=gray!15]
                (2.8,1.6) rectangle (3.6,2.0)
                node[pos=0.5] {$S(D)$};
            \draw[thick] (3.6,1.8) -- (4.1,1.8);

            \draw[thick, fill=white]
                (4.1,1.65) rectangle (4.7,1.95)
                node[pos=0.5, above=0.15cm] {$R_{\mathrm{L}}$};
            \draw[thick] (4.7,1.8) -- (5.0,1.8);

            \draw[thick]
                (5.0,1.8)
                sin (5.2,2.05) cos (5.4,1.8)
                sin (5.6,2.05) cos (5.8,1.8)
                node[above=0.15cm] {$L$}
                -- (6.3,1.8);
            \draw[thick, ->]
                (4.6,1.4) -- (5.4,1.4)
                node[midway, above] {$i_{\mathrm{L}}$};

            \draw[thick] (6.3,1.8) -- (6.3,1.4);
            \draw[thick, fill=white]
                (6.05,1.1) rectangle (6.55,1.4)
                node[pos=0.5, right=0.30cm] {$R_{\mathrm{Co}}$};
            \draw[thick] (6.3,1.1) -- (6.3,0.8);
            \draw[thick] (6.0,0.8) -- (6.6,0.8);
            \draw[thick] (6.0,0.65) -- (6.6,0.65)
                node[right=0.0cm] {$C_{\mathrm{o}}$};
            \draw[thick] (6.3,0.65) -- (6.3,0);

            \draw[thick] (6.3,1.8) -- (8.0,1.8) -- (8.0,1.25);
            \draw[thick, fill=white]
                (8.0,0.95) circle (0.3cm)
                node[left=0.2cm] {$I_{\mathrm{o}}$};
            \draw[thick, ->] (8.0,1.1) -- (8.0,0.8);
            \draw[thick] (8.0,0.65) -- (8.0,0);

            \draw[thick] (0.2,0) -- (8.0,0);
            \filldraw[fill=white] (0.2,0) circle (1.5pt)
                node[left=1pt] {$0\,\mathrm{V}$};

            \draw[thick, ->]
                (1.6,1.5) -- (1.6,0.3)
                node[midway, left] {$v_{\mathrm{Ci}}$};
            \draw[thick, ->]
                (5.7,1.5) -- (5.7,0.3)
                node[midway, left] {$v_{\mathrm{Co}}$};
        \end{scope}

        \begin{scope}[shift={(0.875,-3)}]
            \node[font=\footnotesize\bfseries, anchor=center]
                at (3.225,2.5)
                {(b) Bouc--Wen oscillator};

            \draw[thick] (0,-0.4) -- (0,2.0);
            \foreach \y in {-0.3,0.0,0.3,0.6,0.9,1.2,1.5,1.8}
                \draw[gray] (0,\y) -- (-0.25,\y-0.15);

            \draw[thick, fill=gray!15]
                (4.0,-0.1) rectangle (5.4,1.7)
                node[pos=0.5] {\textbf{Mass} $m_{\mathrm{L}}$};
            \draw[thick, ->]
                (5.4,0.8) -- (6.7,0.8)
                node[midway, above] {$u$};
            \draw[thick, ->]
                (4.7,1.9) -- (5.7,1.9)
                node[midway, above] {$p,\,v$};

            \draw[thick]
                (0,1.6) -- (0.8,1.6)
                sin (1.0,1.8) cos (1.2,1.6)
                sin (1.4,1.4) cos (1.6,1.6)
                sin (1.8,1.8) cos (2.0,1.6)
                sin (2.2,1.4) cos (2.4,1.6)
                node[above] {$k_{\mathrm{L}}$}
                -- (4.0,1.6);

            \draw[thick] (0,0.8) -- (1.2,0.8);
            \draw[thick] (1.2,0.6) -- (1.2,1.0);
            \draw[thick]
                (1.0,0.5) -- (2.0,0.5)
                -- (2.0,1.1) -- (1.0,1.1);
            \draw[thick]
                (2.0,0.8) -- (2.4,0.8)
                node[above] {$c_{\mathrm{L}}$}
                -- (4.0,0.8);

            \draw[thick] (0,0) -- (0.8,0);
            \draw[thick, fill=white]
                (0.8,-0.25) rectangle (2.4,0.25)
                node[pos=0.5] {\textbf{Hysteresis} $z$};
            \draw[thick] (2.4,0) -- (4.0,0);

            \draw[thick] (4.3,-0.22) circle (0.12cm);
            \draw[thick] (5.1,-0.22) circle (0.12cm);
            \draw[thick] (3.8,-0.34) -- (5.6,-0.34);
        \end{scope}

    \end{tikzpicture}%
    }
    \vspace{-0.5cm}
    \caption{Benchmark schematics: (a) equivalent circuit topology of the Buck--Boost converter, and (b) lumped-parameter model of the Bouc--Wen oscillator.}
    \label{fig:sch_benchmarks}
\end{figure}
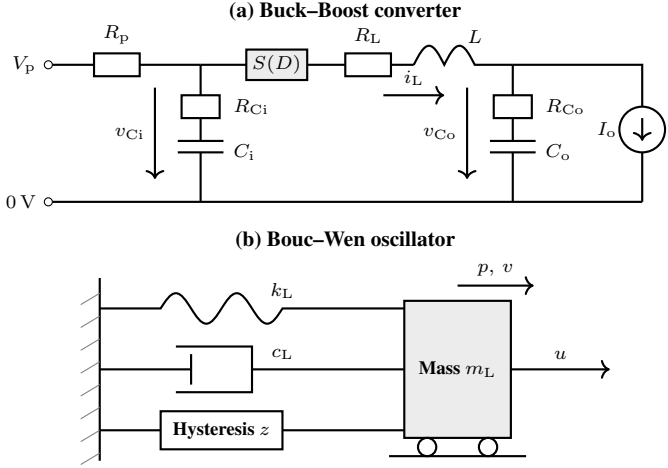

%
The continuous-time benchmark equations are reported below.
In both cases, let $t \in \R_{\geq 0}$ be the time (in $\mathrm{s}$).
\figurename{}~\ref{fig:sch_benchmarks} shows the corresponding schematics.
%
\subsection{Buck--Boost Converter}
\label{ss:app_buck_boost}
%
The continuous-time state vector is:
\begin{equation*}
    \bm{s}(t) \coloneq \ps*{v_{\mathrm{Co}}(t), v_{\mathrm{Ci}}(t), i_{\mathrm{L}}(t)}^\trans \in \R^3.
\end{equation*}
Using the duty-cycle input $u(t)$, the continuous-time dynamics are:
\begin{align*}
    \dot{v}_{\mathrm{Co}}(t) &= \frac{1-u(t)}{C_{\mathrm{o}}}\,i_{\mathrm{L}}(t)-\frac{1}{C_{\mathrm{o}}}\,I_{\mathrm{o}}, \\
    \dot{v}_{\mathrm{Ci}}(t) &= -\frac{1}{C_{\mathrm{i}}R_{\mathrm{t}}}\,v_{\mathrm{Ci}}(t)-\frac{R_{\mathrm{p}}}{C_{\mathrm{i}}R_{\mathrm{t}}}\,i_{\mathrm{L}}(t)+\frac{1}{C_{\mathrm{i}}R_{\mathrm{t}}}\,V_{\mathrm{p}},\\
    \dot{i}_{\mathrm{L}}(t)  &= -\frac{1 - u(t)}{L} \, v_{\mathrm{Co}}(t) + \frac{R_{\mathrm{p}}}{L R_{\mathrm{t}}} \, v_{\mathrm{Ci}}(t) - \frac{R_{\mathrm{eff}}(t)}{L} \, i_{\mathrm{L}}(t) \\
    &\quad + \frac{(1 - u(t)) R_{\mathrm{Co}}}{L} \, I_{\mathrm{o}} + \frac{R_{\mathrm{Ci}}}{L R_{\mathrm{t}}} \, V_{\mathrm{p}},
\end{align*}
where:
\begin{equation*}
    R_{\mathrm{t}} \coloneq R_{\mathrm{p}} + R_{\mathrm{Ci}}, \quad R_{\mathrm{eff}}(t) \coloneq R_{\mathrm{L}} + (1 - u(t)) R_{\mathrm{Co}} + \frac{R_{\mathrm{p}} R_{\mathrm{Ci}}}{R_{\mathrm{t}}}.
\end{equation*}
%
\subsection{Bouc--Wen Oscillator}
\label{ss:app_bouc_wen}
%
The continuous-time state vector is:
\begin{equation*}
    \bm{s}(t) \coloneq \ps*{p(t), v(t), z(t)}^\trans \in \R^3.
\end{equation*}
Using the applied-force input $u(t)$, the continuous-time equations of motion are:
\begin{align*}
    \dot{p}(t) &= v(t), \\
    \dot{v}(t) &= \frac{1}{m_{\mathrm{L}}} \pr*{u(t) - k_{\mathrm{L}} p(t) - c_{\mathrm{L}} v(t) - z(t) },\\
    \dot{z}(t) &= \alpha v(t) - \beta \pr*{ \gamma \abs{v(t)} \, \abs{z(t)}^{\nu-1} z(t) + \delta v(t) \, \abs{z(t)}^\nu }.
\end{align*}
%
\small
\bibliography{bibliography}
\normalsize

\end{document}